\documentclass[twocolumn,showpacs,superscriptaddress,preprintnumbers,amsmath,amssymb]{revtex4-2}
\usepackage{graphicx}
\usepackage{dcolumn}
\usepackage{appendix}
\usepackage{bm}

\usepackage[usenames,dvipsnames]{xcolor}
\usepackage{subfigure}
\usepackage{bbm}
\usepackage{enumerate}
\usepackage[
  bookmarks=true,
  colorlinks,
  linkcolor=blue,
  urlcolor=blue,
  citecolor=blue,
  plainpages=false,
  pdfpagelabels,
  final,
  breaklinks=true
]{hyperref}

\usepackage{titlesec}
\usepackage{array}
\usepackage{dsfont}
\usepackage{physics}
\usepackage{mathpazo}

\newcommand{\bg}[1]{b_{g\vb{#1}}}
\newcommand{\be}[1]{b_{e\vb{#1}}}
\newcommand{\vbn}{\vb{n}}
\newcommand{\vbm}{\vb{m}}
\newcommand{\vbz}{\vb{0}}
\newcommand{\fc}[2]{ \eta_{\vb{#1}\vb{#2}}(q) }
\newcommand{\fcc}[2]{ \eta^*_{\vb{#1}\vb{#2}}(q) }

\begin{document}

\title{High harmonic generation from a Bose-Einstein condensate}

\author{Philipp Stammer}
\email{philipp.stammer@icfo.eu}
\affiliation{ICFO-Institut de Ciencies Fotoniques, The Barcelona Institute of Science and Technology, Castelldefels (Barcelona) 08860, Spain.}
\affiliation{Atominstitut, Technische Universit\"{a}t Wien, 1020 Vienna, Austria}

\date{\today}

\begin{abstract}

Lasers provide intense coherent radiation, essential to cool and trap atoms into a Bose-Einstein condensate or can alternatively drive the non-linear dynamics of high-order harmonic generation. Yet, these two fundamental processes remained of independent consideration. Here, we connect matter waves at ultracold temperatures with radiation bursts on the ultrafast attosecond timescale. We do this by exploring high harmonic generation from a Bose-Einstein condensate.
We show that the quantum state of the generated harmonics of a driven Bose gas is a classical mixture, while below the critical temperature of Bose-Einstein condensation the emitted harmonic radiation is in a pure quantum state. These states furthermore exhibit squeezing and entanglement across all field modes.

\end{abstract}

\maketitle

With the advent of the laser~\cite{schawlow1958infrared} the research in quantum science showed a tremendous leap by manipulating individual quantum systems~\cite{haroche2025laser}. Ranging from the observation of quantum jumps~\cite{bergquist1986observation}, or the ability of trapping ions~\cite{neuhauser1980localized} and cooling atoms~\cite{dalibard1989laser} to the investigation of cavity quantum electrodynamics~\cite{raimond2001manipulating}, just to name a few. Remarkable areas have formed based on the interaction of laser light with atoms. Particularly noteworthy for this work is the cooling of atoms, which below a critical temperature $T_c$ can form a Bose-Einstein condensate (BEC)~\cite{anderson1995observation, davis1995bose}. Furthermore, using intense laser fields allow to study ultrafast dynamics on the attosecond time-scale~\cite{corkum2007attosecond}. The fundamental cornerstone in attosecond science is the process of high-order harmonic generation (HHG)~\cite{lewenstein1994theory}, in which the driving field photons are upconverted to form a frequency comb spanning dozens of harmonic orders~\cite{antoine1996attosecond}. The observation of HHG was followed by the generation of attosecond pulses of radiation~\cite{paul2001observation, hentschel2001attosecond}, paving the way for the field of attosecond physics~\cite{krausz2009attosecond}. 

While the process of HHG has been mostly studied in the semi-classical regime~\cite{lewenstein1994theory}, where the driving laser is assumed to be classical, recent interest in the quantum optical formulation provided a new perspective on the process~\cite{lewenstein2021generation, cruz2024quantum, stammer2023quantum, stammer2025theory, gorlach2020quantum}. For instance, it was shown that quantum optical HHG (QHHG) allows to generate squeezed~\cite{yi2024generation, lange2025excitonic, stammer2024entanglement, rivera2024squeezed, lange2024electron, gorlach2020quantum} or entangled state of light~\cite{yi2024generation, stammer2024entanglement, stammer2022high, theidel2024evidence, rivera2024nonclassical, stammer2022theory}, and that the photon statistics can show bunching~\cite{lemieux2024photon} or anti-bunching signatures~\cite{stammer2025theory}.
In contrast to the interest in QHHG only recently, the full quantum optical description of a Bose gas interacting with light is well studied~\cite{politzer1991light, svistunov1990resonance, lewenstein1994quantum, you1995quantum, you1996quantum, lewenstein1996quantum, moore1999quantum, javanainen1994optical, javanainen1995spectrum, javanainen1995off, zhang1994quantum, morice1995refractive, politzer1997bose, mekhov2012quantum, zhu2016light}. In particular, the quantum optical theory of light scattering from a BEC has shown that the number of emitted photons can dramatically increase below the critical temperature~\cite{lewenstein1993probing}. However, these interactions have focused on relatively low intensity driving fields with single photon or resonant interactions~\cite{cirac1994quantum, ruostekoski1997phase, javanainen1999one, you1994line}, whereas for the highly non-linear process of HHG an intense laser field is crucial, and the involved transitions are non-perturbative.

\begin{figure}
    \centering
	\includegraphics[width = 0.9\columnwidth]{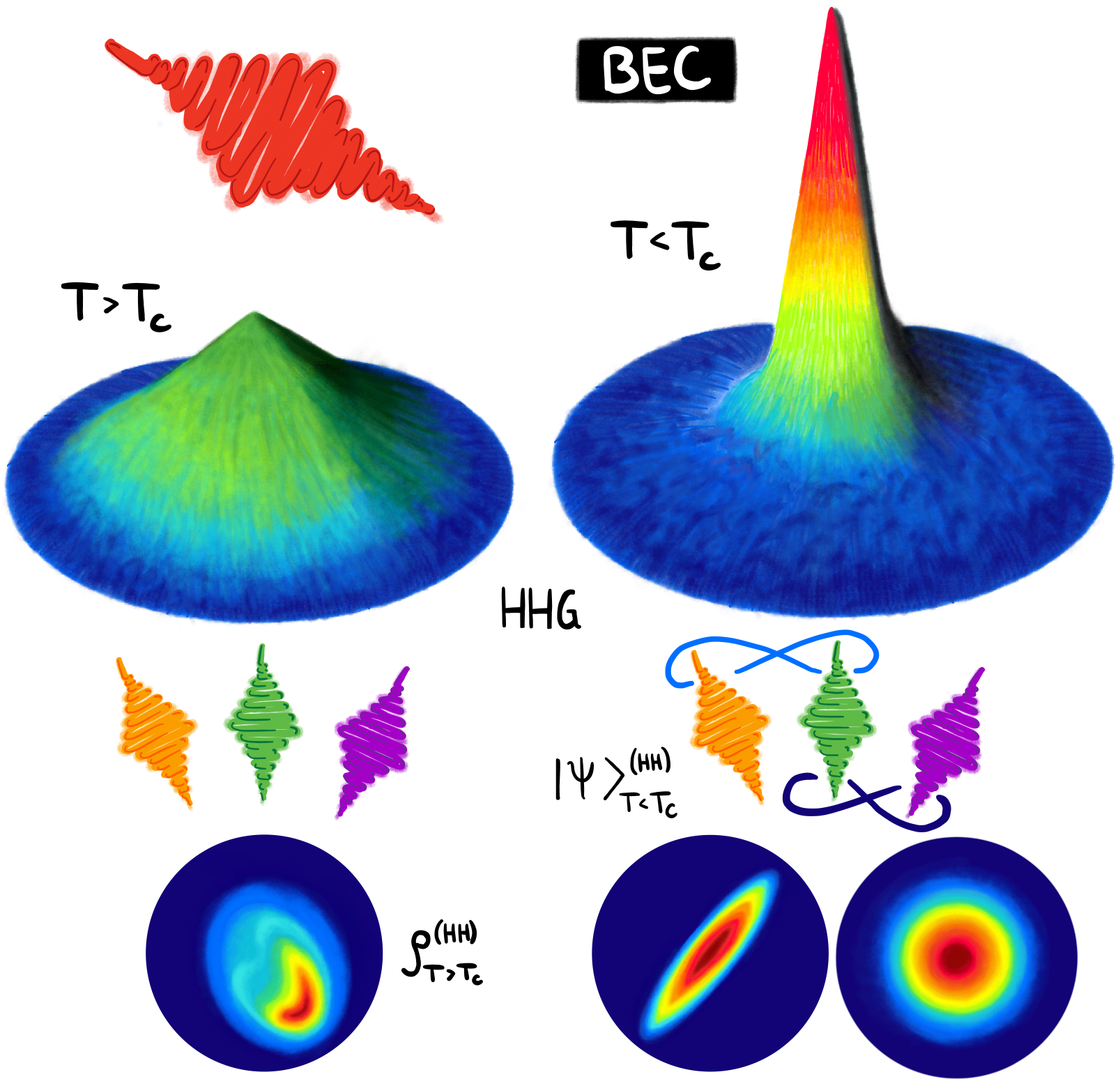}
	\caption{\textbf{HHG from a BEC:} Driving the process of high-harmonic generation (HHG) in a Bose-Einstein condensate (BEC) give rise to distinct quantum states of the harmonics scattered from the condensate. Above the critical temperature ($T > T_c$) the state is a classical mixture of Gaussian states, while below the critical temperature $(T < T_c)$ the harmonic field is in a pure quantum state, exhibiting entanglement and squeezing.}
      \label{fig:cartoon}
\end{figure}

Be it how it may, the two processes of HHG and BEC remained disconnect until now. In this work we resolve this discrepancy by considering the process of HHG emitted from a BEC.   
We are particularly interested in the quantum state of the harmonics, extending existing work on resonant light scattering from a BEC, that so far only focused on the light properties, such as the number of emitted photons~\cite{lewenstein1993probing, javanainen1994optical, javanainen1995spectrum}, but never studied the quantum state of the field itself. 
We show that the quantum state of the harmonics encode the information of the formation of the condensate, revealing that below the critical temperature of Bose-Einstein condensation the emitted harmonics transition from a mixed to a pure quantum state. 
Furthermore, we show that these states can show non-classical signatures by virtue of entanglement and squeezing across all field modes.
From a different perspective, we propose a scheme for probing Bose-Einstein condensation using ultrashort and intense laser pulses via the processes of high-order harmonic generation.
With this we ultimately connect the realm of ultracold matter waves with the domain of ultrafast non-linear light generation.

\section*{Framework}

\begin{figure}
    \centering
	\includegraphics[width = 0.55\columnwidth]{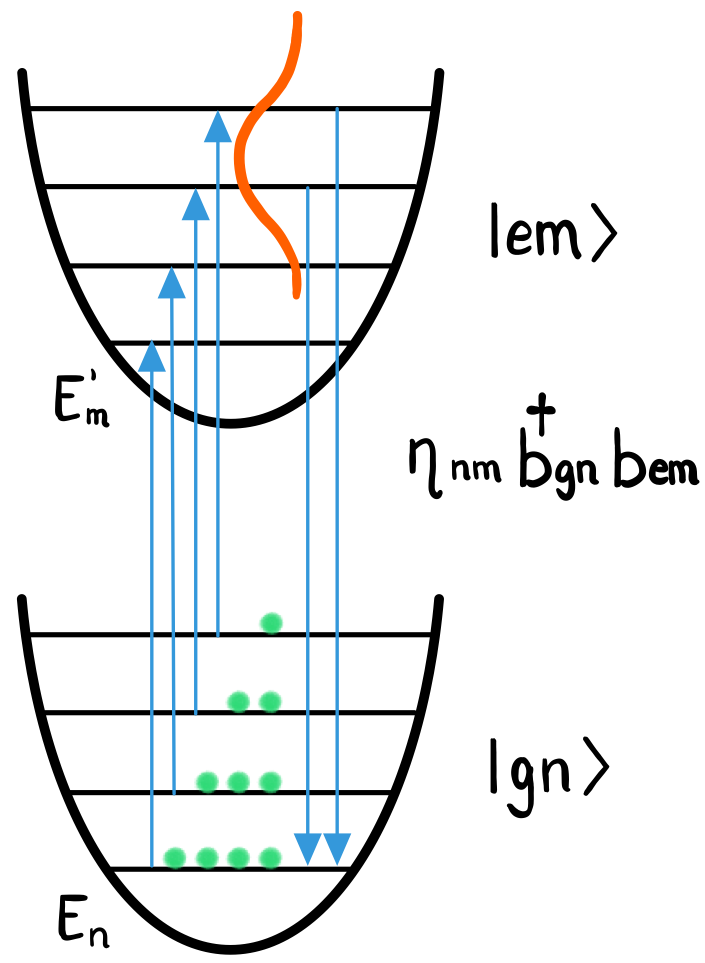}
	\caption{\textbf{Bosonic modes:} Schematic representation of the mode structure of the driven Bose gas. The atoms in the internal electronic ground state $\ket{g}$ are trapped in a harmonic oscillator potential with the center of mass modes $\ket{g\vb{n}} \equiv \ket{\vbn}$ of energy $E_{\vb{n}}$. Due to the interaction with the short and intense driving pulse, the electronic degree of freedom is excited into a continuum state $\ket{e}$, such that the corresponding atoms are associated to the excited state potential $\ket{e\vb{m}} \equiv \ket{\vbm}$ with energy $E_{\vb{m}}'$. The driving field induces transitions from each ground state atom into a wavepacket (orange) of excited state atoms governed by the Franck-Condon like factors $\eta_{\vb{n}\vb{m}}$.}
      \label{fig:level_scheme}
\end{figure}

To describe the interaction between an ideal Bose gas with an intense and pulsed light field, we follow the full quantum optical approach of many-body systems given by the quantum field theory of atoms interacting with photons~\cite{lewenstein1994quantum, you1995quantum}. However, we go beyond these approaches since we do not employ the rotating-wave approximation and further consider the specific characteristics of the dynamics underlying HHG~\cite{lewenstein1994theory}. In the following, we will see that the harmonic quantum state depends on the Bose-Einstein statistics of the trapped atoms. 
We consider the atoms to be trapped in an isotropic harmonic oscillator potential of frequency $\omega_T$, such that he eigenstates of the center of mass motion of the atoms in the trap $\ket{\vbn}$ are labeled by $\vbn = (n_x, n_y, n_z)$, with the corresponding energy $E_{\vbn} = \omega_T (n_x + n_y + n_z)$.
Consequently, the index $\vbz \equiv (0_x, 0_y, 0_z)$ is the lowest energy state in the trap. 
The low-frequency intense driving laser field, crucial for the highly non-linear process of HHG far from resonance, induces transitions between the internal atomic ground and excited state ($\ket{g} \to \ket{e}$). 
For the process of HHG these excited internal states are in fact states of an ionized electron, in which the electron tunnel ionizes to continuum states, before recombining back to the internal ground state by emitting high harmonic photons. The energy cap between the internal states is therefore given by the kinetic energy of the electron in the continuum.  
Since the ideal Bose gas interacts with an intense and pulsed laser field, represented by a coherent state $\ket{\alpha}$, the induced transitions between the internal ground and excited state are given by a dipole interaction with the intense classical field $E_{cl}(t) \propto \abs{\alpha} \sin(\omega_L t)$ of frequency $\omega_L$.
We consider the ideal (non-interacting) Bose gas with Hamiltonian $H_{BG}$ and the electromagnetic field of modes $q$ and frequencies $\omega_q = q \omega_L$;
\begin{align}
    H_{BG} & = \sum_{\vbn}  E_{\vbn} \bg{n}^\dagger \bg{n} + \sum_{\vbm} E_{\vbm}^\prime \be{m}^\dagger \be{m}, \\
    H_F & = \sum_q \omega_q a_q^\dagger a_q,
\end{align}
where the atomic operator $\bg{n}^{(\dagger)}$ denotes annihilation (creation) of an excitation in the $n$-th state of the ground state potential in the rotational invariant trap. The atomic operators $\be{m}^{(\dagger)}$ denotes annihilation (creation) of an excitation in the excited state potential of energy $E_{\vbm}^\prime$, which are shifted from the ground state energy by the energy of the excited electron. The energy difference is the energy of the atom gained by the electron due to the driving field. Within the strong field approximation (SFA) this energy gain is given by the kinetic energy of the electron after ionization~\cite{amini2019symphony}.
The bosonic operators of the field are given by $a_{q}^{(\dagger)}$, and all introduced operators fulfill standard bosonic commutation relations. A schematic representation of the mode structure of the Bose gas is given in Fig.~\ref{fig:level_scheme}.

A key difference to previous work about resonant excitation into internal excited atomic states~\cite{lewenstein1993probing, javanainen1994optical, javanainen1995spectrum}, is the strong field induced dynamics of the electron~\cite{lewenstein1994theory}.  
The interaction Hamiltonian describes the process in which an electron transitions via tunnel ionization (instantaneously), driven by the intense laser field, from the atoms in the internal ground state into the excited state potential
\begin{align}
    H_I (t) = - i  \sum_{\vb{n}, \vb{m}, q} & \xi(q) \left[ \fc{n}{m} \bg{n}^\dagger \be{m} + \fcc{n}{m} \be{m}^\dagger \bg{n} \right] \nonumber \\
    & \times \left[ a_q^\dagger e^{i \omega_q t} - a_q e^{- i \omega_q t}\right],
\end{align}
where the Franck-Condon like factors are the matrix elements for the transitions from the $n$-th state of the ground state potential to the $m$-th state of the excited state potential $\fc{n}{m} = \bra{\vbn} e^{- i \vb{k}_q\cdot \vb{r}} \ket{\vbm}$, and the Hamiltonian is given in the rotating frame with respect to $H_F$. 
This atom-photon interaction Hamiltonian includes, for instance, the first term which is the process when an atom in the internal excited state $\ket{\vbm}$ transitions to the internal ground state corresponding to the center of mass state $\ket{\vbn}$ weighted by the matrix element $\xi(q) \eta_{\vbn \vbm}(q)$, with the light-matter coupling constant $\xi(q) \propto \sqrt{\omega_q}$. This dipole transition is due to the coupling with the electric field operator of all modes $E_Q(t) = - i \sum_q \xi(q) [a_q^\dagger e^{i \omega_q t} - a_q e^{- i \omega_q t}] $. 
Note that this interaction Hamiltonian goes beyond the common assumptions performed in Refs.~\cite{lewenstein1994quantum, you1995quantum, lewenstein1993probing, javanainen1994optical}, where a two-level system is driven by a resonant interaction in the rotating-wave approximation. Going beyond the two-level approximation allows to consider the case of strong field induced non-linear effects, such as the process of HHG.
Here, we emphasize that the light-matter interaction Hamiltonian in the length and velocity gauge are in principle unitarily equivalent, although performing a two-level approximation breaks this unitary equivalence~\cite{lewenstein1994quantum}. However, since we do not perform this approximation, the interaction Hamiltonian is gauge invariant. 
Furthermore, considering that we drive the Bose gas by the external intense laser field, we add the semi-classical interaction to the Hamiltonian
\begin{align}
    H_{I,cl} (t) = 2 \abs{\alpha} \, \xi(k_L) \sum_{\vbn, \vbm} & \sin(\omega_L t)  \left[ \eta_{\vbn \vbm}(k_L) \, \bg{\vbn}^\dagger \be{\vbm} \right. \\
    & \left. + \, \eta^*_{\vbn \vbm}(k_L) \, \be{\vbm}^\dagger \bg{\vbm} \right]. \nonumber
\end{align}
where $k_L$ is the wavevector of the driving field of frequency $\omega_L$.

\section*{HHG from a BEC}

Now, the central question of this paper is the structure of the quantum optical state of the harmonic field modes after the strong field interaction with the ultracold Bose gas. 
We therefore assume that the Bose gas initially populates the ground state potential energy levels according to the Bose-Einstein distribution, so that the number of atoms in the $n$-th state is $N_{\vbn} = z \, e^{- \beta E_{\vbn}}/(1- z \, e^{- \beta E_{\vbn}})$, where $\beta = 1/T$ is the inverse temperature and $z$ is the fugacity. Considering that we have $N$ atoms in the trap, the constraint of conserved particle number is $\sum_{\vbn} N_{\vbn} = N$. Remember that below the critical temperature of Bose-Einstein condensation $T_c$, the number of atoms in the lowest energy state $N_{\vbz}$ becomes extensive and $z=1$~\cite{bagnato1987bose}. 
Consequently, for the total initial state we have the mixture $\rho(0) = \sum_{\vbn} P(\vbn) \dyad*{\{ 0_q\}, \vbn, \vbz }$, with $P(\vbn)$ the occupation probability of the atom distribution $\vbn$ in the ground state potential. The harmonic modes of the electromagnetic field are initially in the vacuum $\ket*{\{ 0_q \}} = \bigotimes_q \ket*{0_q}$. Finally, the Bose gas follows the aforementioned Bose-Einstein distribution in the ground state potential $\rho_{BG} = \sum_{\vbn} P(\vbn) \dyad{\vbn}$ where the electron is in the internal ground state, and the potential corresponding to the excited electronic state is initially unpopulated $\ket{ \vbz } = \bigotimes_{\vbm} \ket{ 0_{\vbm} }$.

To solve the dynamics of the quantum state of the optical field modes, we further transform into the interaction picture with respect to $H_{sc}(t) = H_{BG} + H_{I,cl}(t)$, such that the final interaction Hamiltonian is given by
\begin{align}
\label{eq:interaction_hamiltonian}
    H_{I,Q}(t) = - i \sum_q \xi(q) D_q(t) \left[ a_q^\dagger e^{i \omega_q t} - a_q e^{- i \omega_q t} \right].
\end{align}

The time-dependent dipole $D_q(t) = U_{sc}^\dagger(t) D_q U_{sc}(t)$ is the dipole moment operator
\begin{align}
    D_q = \sum_{\vbn, \vbm}   \left[ \fc{\vbn}{\vbm} \, \bg{\vbn}^\dagger \be{\vbm} + \fcc{\vbn}{\vbm} \, \be{\vbm}^\dagger \bg{\vbn} \right],
\end{align}
dressed by the semi-classical interaction, where $U_{sc}(t)$ is the unitary rotation with respect to the semi-classical Hamiltonian $H_{sc}(t)$.
This transformation is performed to solve the dominant semi-classical dynamics from the intense driving field self-consistently, and treat the driven dipole as the source term coupled to the quantum field. 
We shall now proceed to solve the dynamics of the optical field modes. Therefore, we consider each term in the classical mixture of the initial state $\rho(0) = \sum_{\vbn} P(\vbn) \dyad{\psi_{\vbn}(0)}$ separately.
When solving the Schrödinger equation for the harmonic quantum states we are interested in the case where the atoms return to the ground state potential~\cite{lewenstein2021generation, stammer2023quantum}, inspired by the 3-step model underlying HHG~\footnote{The 3-step model of HHG~\cite{corkum1993plasma} provides an intuitive picture of the underlying electron dynamics. Due to the intense laser field, the electron can tunnel (instantaneously) into the continuum through the barrier formed by the atomic binding potential and the dipole interaction with the driving field. This is followed by an acceleration of the electron in the continuum due to the presence of the intense field, which finally brings the electron back to the core where it can recombine to its ground state by emitting high-order harmonic radiation.}. 
The dynamics of the field modes are thus given by (for details see End Matter~\ref{end:derivation_TDSE})
\begin{align}
\label{eq:coupled_DGL}
    i \partial_t & \ket{\Phi_{\vbn, \vbz}(t)} = \bra{\vbn, \vbz} H_{I,Q}(t) \ket{\vbn, \vbz} \ket{\Phi_{\vbn, \vbz}(t)} \\
    & + \sum_{\vbn^\prime \neq \vbn} \sum_{\vbm} \bra*{\vbn, \vbz} H_{I,Q}(t) \ket*{\vbn^\prime, \vbm} \ket*{\Phi_{\vbn^\prime, \vbm}(t)}, \nonumber
\end{align}
where we have defined the conditioned photonic quantum state $\ket{\Phi_{\vbn, \vbz}(t)} \equiv \bra{\vbn, \vbz} \ket{\psi(t)}$, and introduced an identity on the Hilbert space of the Bose gas $\mathds{1} = \sum_{\vbn} \dyad{\vbn} \otimes \sum_{\vbm} \dyad{\vbm}$.
The right hand side has an intuitive physical interpretation: the first term corresponds to the mean value of the dipole, $\bra{\vbn, \vbz} D(t) \ket{\vbn, \vbz}$, when all atoms are in the internal ground state distributed along $\vbn$. While the second term corresponds to the transitions from the excited internal states into the ground state $\ket{\vbn^\prime, \vbm} \to \ket{\vbn, \vbz}$.
Due to the particular strong field dynamics we can neglect the contribution from the states in which $\vbm \neq \vbz$. This is because $|| \ket*{\Phi_{\vbn, \vbm \neq \vbz} (t)} || \ll || \ket*{\Phi_{\vbn,\vbz} (t)} ||$, following the strong field approximation~\cite{amini2019symphony}. 
Within the SFA we can further assume that the short laser pulse only influences the electron dynamics, such that the interaction leaves the center of mass motion unaffected. Hence, it remains to solve 
\begin{align}
\label{eq:TDSE_conditioned}
    i \partial_t \ket{\Phi_{\vbn, \vbz} (t)} = \left[ H_{I,Q}(t) \right]_{\vbn \, \vbz, \vbn \, \vbz} \ket{\Phi_{\vbn, \vbz}(t)},
\end{align}
where 
\begin{align}
\label{eq:hamiltonian_effective}
    \left[ H_{I,Q}(t) \right]_{\vbn \, \vbz, \vbn \, \vbz} = \sum_q \bra{\vbn, \vbz} D_q(t) \ket{\vbn, \vbz} E_q(t),
\end{align}
is the effective Hamiltonian with the electric field operator of each mode $E_q(t) = - i \xi(q) ( a_q^\dagger e^{i \omega_q t} - a_q e^{- i \omega_q t})$.
Since the commutator at different times of the effective Hamiltonian acting on the field in Eq.~\eqref{eq:hamiltonian_effective} is a c-number, the solution to \eqref{eq:TDSE_conditioned} is given by 
\begin{align}
    \ket{\Phi_{\vbn, \vbz}} = \prod_q D[\chi_q(\vbn)] \ket{\{ 0_q \}}= \bigotimes_q \ket{\chi_q(\vbn)},
\end{align}
with the coherent state amplitudes 
\begin{align}
    \chi_q(\vbn) = \xi(q) \int_{-\infty}^\infty dt \ e^{i \omega_q t} \bra{\vbn, \vbz} D_q(t) \ket{\vbn, \vbz}.
\end{align}

And consequently, the final state of the high harmonics emitted from the driven Bose gas above the critical temperature is a classical mixture over product coherent states 
\begin{align}
\label{eq:solution_above}
    \rho_{T > T_c}^{(HH)} = \sum_{\vbn} P(\vbn) \bigotimes_q \dyad{\chi_q(\vbn)}.
\end{align}

Now, the situation changes dramatically below the critical temperature $T< T_c$, for which the solution of the harmonic quantum state can be easily inferred. We now find the Bose gas in the condensate, where we assume that $N_{\vbz}$ atoms constitute a coherent wavepacket and occupy the lowest energy state $\vbn = (0_x, 0_y, 0_z)$. 
Due to the macroscopic occupation of the BEC in the lowest energy state, we only keep the condensate mode $\bg{\vbz}$ for the ground state potential and the excited states $\be{\vbm}$, since (almost) all atoms occupy the lowest energy state~\cite{javanainen1994optical}. It thus directly follows from \eqref{eq:solution_above} that the harmonics emitted from a BEC are given by a pure quantum state 
\begin{align}
\label{eq:solution_below}
    \rho_{T< T_c}^{(HH)} = \bigotimes_q \dyad{\chi_q(\vbn = \vbz)}.
\end{align}

This shows that for the Hamiltonian under the strong field approximation of Eq.~\eqref{eq:hamiltonian_effective}, the final quantum optical state above $T_c$ is a mixture, while below the critical temperature we have a pure state.
However, it remains to consider the scenario in which the population of the excited state potential is not negligible, and thus going beyond the linearizaton of the effective dynamics. We thus go back to Eq.~\eqref{eq:coupled_DGL}, and explicitly take into account the second term in which $\vbm \neq \vbz$. We thus need to find $\ket*{\Phi_{\vbn^\prime, \vbm}(t)}$, which we obtain by solving the corresponding conditioned Schrödinger equation. Since we are interested in the first order correction of the dynamics due to the excited state population to the coherent state solution, and the transition matrix elements $\bra{\vbn, \vbz} H_{I,Q}(t) \ket{\vbn^\prime, \vbm}$ are already in first order of $\vbm$, we need the zero-th order correction in the solution to $\ket*{\Phi_{\vbn^\prime, \vbm}(t)}$. Therefore, we solve the coupled equations of $\ket{\Phi_{\vbn, \vbz}(t)}$ and $\ket{\Phi_{\vbn^\prime, \vbm}(t)}$ from \eqref{eq:coupled_DGL} up to first order in the correction of the transition matrix elements. 
Doing so, we obtain for the final field state below the critical temperature $T < T_c$ (see End Matter~\ref{end:squeezing} for a detailed derivation)
\begin{align}
\label{eq:solution_squeezing}
    \ket{\zeta(\vbn = \vbz)}_{T < T_c} = \mathcal{S}_{MS}[\zeta(\vbz)] \ket{ \{ 0_q \}},
\end{align}
which is a multi-mode squeezed vacuum state of all harmonics. The multi-mode squeezing operator is given by 
\begin{align}
    \mathcal{S}_{MS}[\zeta(\vbn)] = \exp \left[ \sum_{q,p} \left( \zeta_{q,p}^{(\vbn)} \, a_q^\dagger a_p^\dagger  - \zeta_{q,-p}^{(\vbn)} \, a_q^\dagger a_p + \operatorname{H.c.} \right) \right],
\end{align}
with the squeezing parameter obeying $\zeta_{q,p}^{(\vbn)} = (\zeta_{-q,-p}^{(\vbn)})^* $, and is given by 
\begin{align}
\label{eq:squeezing_parameter}
    \zeta_{q,p}^{(\vbn)} = & \xi(q) \xi(p) \int_0^\infty dt \int_0^t d\tau \sum_{\vbn^\prime \neq \vbn} \sum_{\vbm \neq \vbz} \ e^{i \omega_q t} \ e^{i \omega_p \tau} \\
    & \times \bra{\vbn, \vbz} D_q(t) \dyad*{\vbn^\prime, \vbm} D_p(\tau) \ket{\vbn, \vbz}. \nonumber
\end{align}

We can see that the correlations of the dipole moment are responsible for the appearance of entanglement and squeezing in the harmonic field modes~\cite{stammer2024entanglement, lange2024electron}.
Note that alike the case of negligible excited state population in Eq.~\eqref{eq:solution_above}, the final quantum optical state above the critical temperature $T > T_c$ would also correspond to a classical mixture of different squeezed stats $\ket{\zeta(\vbn)}$ following the initial distribution $P(\vbn)$. Such a mixture destroys the quantum squeezing properties of the total state, such that above the critical temperature the state has no genuine quantum squeezing below the shot noise limit. 
We recall that the central question of this paper was to find the structure of the quantum optical state of the generated harmonics from a Bose-Einstein condensate, which we have answered by virtue of Eq.~\eqref{eq:solution_below} and \eqref{eq:solution_squeezing}. However, this leads, of course, to many other questions. For instance, the particular influence of the statistics of the ultracold bosons on light scattering~\cite{lu2023bosonic}. Likewise, the amplitudes $\chi_q(\vbn)$ and $\zeta_{q,p}(\vbn)$ of the harmonic field modes remain to be calculated, for which a Heisenberg picture approach similar to Ref.~\cite{lewenstein1993probing} is more suitable and will be discussed elsewhere. 

The observation of this paper can be seen as a powerful picture, namely, that not only the Bose gas condense into a BEC but also that the state of the emitted harmonics "condense" from a classical mixture into a pure quantum state below the critical temperature.

\section*{Discussion}

In this work we have studied an ideal Bose gas probed by an intense and short laser pulse, and obtained the quantum optical state of the scattered field modes after the process of high-order harmonic generation. We have seen that the quantum state of the harmonics is given by a classical mixture for the Bose gas above the critical temperature $T_c$. However, below $T_c$ the harmonic field modes emitted from a Bose-Einstein condensate transitions into a pure quantum state, exhibiting non-classical characteristics by virtue of entanglement and squeezing.

Continuing on this work, we envision that the investigation of intense light fields interacting with many-body quantum systems, such as ultracold Bose gases, is only at the beginning to present its full potential. Extending the analysis to interacting Bose gases~\cite{konstantinou2025interacting}, the generation of massive entangled light states due to the BEC interaction~\cite{ng2009entangled} or the study of entanglement between the different bosonic field degrees of freedom leading to hybrid man-body entanglement~\cite{sorensen2001many}, are only some of the possible directions. We note, that beyond Bose gases, extension towards high harmonic light scattering from degenerate Fermi gases seems an interesting direction~\cite{deb2021observation, margalit2021pauli, sanner2021pauli}, in which the Pauli blocking leads to a suppression in the photon scattering. 
Furthermore, considering ionization of ultracold quantum gases into a plasma~\cite{killian1999creation, ciampini2002photoionization, mazets1998photoionization} could be extended towards strong field above-threshold ionization~\cite{wessels2018absolute, rivera2022light}, and is therefore complementary to the investigations of using analog quantum simulation of strong field driven processes~\cite{arlinghaus2010driven, sala2017ultracold, senaratne2018quantum, arguello2025quantum, arguello2024analog}.

In summary, this work has established the connection of strong laser field physics with ultracold atoms, i.e. it connects quantum optical HHG with ultracold quantum gases, specifically Bose-Einstein condensation. 
Since Bose-Einstein condensates exhibit strong quantum many-body effects, this can lead to unique HHG phenomena absent in simpler systems, such that the rich structure of Bose gases presents a new platform to study many-body effects on the emission of high-order harmonics~\cite{silva2018high}.
Generating high-order harmonics from a BEC opens up the possibility of exploring novel regimes of HHG, potentially leading to new applications in attosecond science, or to measure the dynamics of an ultracold quantum gas on the attosecond time-scale.

\begin{acknowledgments}
\textit{I dedicate this paper to my PhD supervisor Maciej Lewenstein, to honor the significant contributions he made in these two fields. This paper was prepared for the occasion of the 70th birthday of Maciej.} \\

P.S. acknowledges funding from the European Union’s Horizon 2020 research and innovation programe under the Marie Skłodowska-Curie grant agreement No 847517.
ICFO acknowledges support from: European Research Council AdG NOQIA; MCIN/AEI (PGC2018-0910.13039/501100011033,  CEX2019-000910-S/10.13039/501100011033, Plan National STAMEENA PID2022-139099NB, project funded by MCIN/AEI/10.13039/501100011033 and by the “European Union NextGenerationEU/PRTR" (PRTR-C17.I1), FPI); project funded by the EU Horizon 2020 FET-OPEN OPTOlogic, Grant No 899794, QU-ATTO, 101168628), Fundació Cellex; Fundació Mir-Puig; Generalitat de Catalunya (European Social Fund FEDER and CERCA program.

\end{acknowledgments}

\bibliography{literatur}{}

\appendix
\section*{Appendix}

\section{\label{end:derivation_TDSE}Derivation of the conditioned Schrödinger equation}

The central question of this paper is the quantum state of the harmonic field modes, which is obtained when evolving the initial state $\rho(0)$ by the Hamiltonian $H_{I,Q}(t)$ (see Eq.~\eqref{eq:interaction_hamiltonian} of the main text), and given by $\rho(t) = \sum_{\vbn} P(\vbn) \dyad{\psi_{\vbn}(t)}$, where $\ket{\psi_{\vbn}(t)} = U(t) \ket{\psi_{\vbn}(0)}$.
Since we are interested in the case of HHG, we condition on the case where all atoms are in the internal ground state, and assume that the atoms are distributed according to $\ket{\vbn^\prime}$ after the interaction. The conditioned HHG state is therefore given by 
\begin{align}
    \bra*{\vbn^\prime, \vbz} \rho(t) \ket*{\vbn^\prime, \vbz} = \sum_{\vbn} P(\vbn) \dyad*{\Phi^{(\vbn)}_{\vbn^\prime, \vbz}(t) },
\end{align}
and it remains to solve for $\ket*{\Phi^{(\vbn)}_{\vbn^\prime, \vbz}(t) } = \bra*{\vbn^\prime, \vbz}\ket*{\Psi_{\vbn}(t)}$. 
With the Schrödinger equation for $\ket*{\Phi^{(\vbn)}_{\vbn^\prime, \vbz}(t) }$ we can use Eq.~\eqref{eq:TDSE_conditioned} such that we obtain 
\begin{align}
    \ket*{\Phi^{(\vbn)}_{\vbn^\prime, \vbz}(t) } = \prod_q D[\chi_q(\vbn^\prime)] \ket*{\Phi^{(\vbn)}_{\vbn^\prime, \vbz}(0) }.
\end{align}

Since the initial state is given by $\ket*{\psi_{\vbn}(0)} = \ket*{\{ 0_q \}, \vbn, \vbz}$ we have 
\begin{align}
    \ket*{\Phi^{(\vbn)}_{\vbn^\prime, \vbz}(0) } = \bra*{\vbn^\prime, \vbz} \ket{\psi_{\vbn}(0)} = \begin{cases}
        0, & \vbn^\prime \neq \vbn \\
        \ket{\{ 0_q \} }, &  \vbn^\prime = \vbn.
    \end{cases}
\end{align}

Taking into account that the atoms can return to any configuration $\ket{\vbn^\prime}$, we add the different possibilities incoherently 
\begin{align}
    \sum_{\vbn^\prime} \bra*{\vbn^\prime, \vbz} & \rho(t) \ket*{\vbn^\prime, \vbz} = \sum_{\vbn^\prime, \vbn} P(\vbn)  \dyad*{\Phi^{(\vbn)}_{\vbn^\prime, \vbz}(t) } \nonumber \\
    & = \sum_{\vbn} P(\vbn) \bigotimes_q \dyad{\chi_q(\vbn)},
\end{align}
which coincides with Eq.~\eqref{eq:solution_above} from the main text, and where we have used that $\ket*{\Phi^{(\vbn)}_{\vbn^\prime, \vbz}(t) } = \delta_{\vbn^\prime, \vbn} \bigotimes_q \ket*{\chi_q(\vbn^\prime)}$.

\section{\label{end:squeezing}Derivation of the squeezing term}

To solve the coupled equations for $\ket*{\Phi_{\vbn, \vbz}(t)}$ and $\ket*{\Phi_{\vbn^\prime, \vbm}(t)}$, we have in addition to Eq.~\eqref{eq:coupled_DGL} the following Schrödinger equation
\begin{align}
\label{eq:TDSE_endmatter}
    i \partial_t \ket*{\Phi_{\vbn^\prime, \vbm}(t)} = \sum_{\vbn^{\prime \prime}} \sum_{\vbm^\prime} \bra*{\vbn^\prime, \vbm} H_{I}(t) \ket*{\vbn^{\prime \prime}, \vbm^\prime}  \ket*{\Phi_{\vbn^{\prime\prime}, \vbm^\prime}(t)}.
\end{align}

Following the arguments of the main text that we are only interested in the zero-th order contribution of $\ket*{\Phi_{\vbn^\prime, \vbm}(t)}$, while $\vbm \neq \vbz$, we only keep the terms in which $\vbn^{\prime \prime} = \vbn$ and $\vbm^\prime = \vbz$.
It thus remains to solve
\begin{align}
    i \partial_t \ket*{\Phi_{\vbn^\prime, \vbm}(t)} =  \bra*{\vbn^\prime, \vbm} H_{I,Q}(t) \ket*{\vbn, \vbz} \ket*{\Phi_{\vbn, \vbz}(t)},
\end{align}
which we can formally integrate (using $\ket*{\Phi_{\vbn^\prime, \vbm}(0)} = 0$)
\begin{align}
    \ket*{\Phi_{\vbn^\prime, \vbm}(t)} = - i \int_0^t d\tau \left[ H_{I,Q}(\tau) \right]_{\vbn^\prime \vbm, \vbn \vbz} \ket{\Phi_{\vbn, \vbz}(\tau)}.
\end{align}

Plugging this into Eq.~\eqref{eq:coupled_DGL}, and considering a Markov type approximation $\ket{\Phi_{\vbn, \vbz}(\tau)} \to \ket{\Phi_{\vbn, \vbz}(t)}$ (see Refs.~\cite{stammer2024entanglement, lange2024hierarchy} for a detailed discussion on its validity), we obtain 
\begin{align}
\label{eq:tdse_2ndorder}
    i \partial_t \ket{\Phi_{\vbn, \vbz}(t)} = \left\{ \left[ H_{I}(t) \right]_{\vbn \vbz, \vbn \vbz} + \left[ \mathcal{H}(t)\right]_{\vbn \vbz, \vbn \vbz} \right\} \ket{\Phi_{\vbn, \vbz}(t)},
\end{align}
where the second-order quantum optical correction term is given by 
\begin{align}
    \left[ \mathcal{H}(t)\right]_{\vbn \vbz, \vbn \vbz} = &  - i \int_0^t d\tau \sum_{\vbn^\prime \neq \vbn} \sum_{\vbm \neq \vbz} \bra*{\vbn, \vbz} H_I(t) \ket*{\vbn^\prime, \vbm} \nonumber \\
    & \times \bra*{\vbn^\prime, \vbm} H_I(\tau) \ket*{\vbn, \vbz}.
\end{align}

Recalling that the Hamiltonian $H_I(t)$ is linear in the field operators, we find that $\left[ \mathcal{H}(t)\right]_{\vbn \vbz, \vbn \vbz}$ is of quadratic order, leading to squeezing signatures. Now solving Eq.~\eqref{eq:tdse_2ndorder}, and using Eq.~\eqref{eq:interaction_hamiltonian} we find for the 2nd order quantum electrodynamical correction that 
\begin{align}
    \int_0^\infty dt \left[ \mathcal{H}(t)\right]_{\vbn \vbz, \vbn \vbz} = i\sum_{q,p} \left( \zeta_{q,p}^{(\vbn)} \, a_q^\dagger a_p^\dagger  - \zeta_{q,-p}^{(\vbn)} \, a_q^\dagger a_p + \operatorname{H.c.} \right),
\end{align}
where we have defined the squeezing parameter 
\begin{align}
    \zeta_{q,p}^{(\vbn)} = & \xi(q) \xi(p) \int_0^\infty dt \int_0^t d\tau \sum_{\vbn^\prime \neq \vbn} \sum_{\vbm \neq \vbz} \ e^{i \omega_q t} \ e^{i \omega_p \tau} \\
    & \times \bra*{\vbn, \vbz} D_q(t) \dyad*{\vbn^\prime, \vbm} D_p(\tau) \ket*{\vbn, \vbz}. \nonumber
\end{align}

With this, the final solution to the field state including the population of the internal excited state, is given by a squeezed and entangled state over all harmonic modes~\cite{stammer2024entanglement}, shown in the frame displaced by the coherent amplitude $\chi_q(\vbn)$;
\begin{align}
    \rho_{T > T_c} = \sum_{\vbn} P(\vbn) \dyad{\zeta(\vbn)},
\end{align}
where $\ket{\zeta(\vbn)}$ is a multi-mode squeezed vacuum state across all harmonic field modes 
\begin{align}
    \ket{\zeta(\vbn)} = \mathcal{S}_{MS}[\zeta(\vbn)] \ket{ \{ 0_q \} }.
\end{align}

\end{document}